\documentclass[twocolumn,english]{revtex4-1}
\usepackage[T1]{fontenc}
\usepackage[latin9]{inputenc}
\usepackage[a4paper]{geometry}
\usepackage{amsmath}

\makeatletter
\usepackage{hyperref}

\makeatother

\usepackage{babel}
\begin{document}
\title{Multiparticle solutions to Einstein's equations}
\author{Humberto Gomez$^{a,b}$}
\author{Renann Lipinski Jusinskas$^{c}$}
\affiliation{$^{a}$ Department of Mathematical Sciences, Durham University, ~\\
Stockton Road, DH1 3LE Durham, UK}
\affiliation{$^{b}$ Facultad de Ciencias Basicas, Universidad Santiago de Cali,~\\
Calle 5 $N^\circ$ 62-00 Barrio Pampalinda, Cali, Valle, Colombia}
\affiliation{$^{c}$ Institute of Physics of the Czech Academy of Sciences \& CEICO
~\\
Na Slovance 2, 18221 Prague, Czech Republic}
\begin{abstract}
In this letter we present the first multiparticle solutions to Einstein's
field equations in the presence of matter. These solutions are iteratively
obtained via the perturbiner method, which can circumvent gravity's
infinite number of vertices with the definition of a multiparticle expansion
for the inverse spacetime metric as well. Our construction provides
a simple layout for the computation of tree level field theory amplitudes
in $D$ spacetime dimensions involving any number of gravitons and
matter fields, with or without supersymmetry.
\end{abstract}
\maketitle

\section{Overview}

Gravity is still in many ways the least understood of the fundamental
forces of nature, arguably at the macroscopic but definitely at the
microscopic level. The former is splendidly described by the general
theory of relativity, while the latter hopefully tangible by the long
sought for theory of quantum gravity.

As a first approximation, the Einstein-Hilbert action can be seen
as a common denominator in this range of scales. It yields as classical
equations of motion Einstein's field equations, and offers a natural
path for a (quantum) field theory of gravitons, the messengers of
gravity.

From this field theory perspective, gravity contains an infinite number
of vertices and is, in fact, non-renormalizable. Even at tree level,
the computation of graviton scattering amplitudes quickly becomes
impractical using standard Feynman diagrams (e.g. \cite{Grisaru:1975ei}).

Modern scattering-amplitude techniques have overcome this problem.
Among them the Britto-Cachazo-Feng-Witten (BCFW) recursion \cite{Britto:2005fq,Benincasa:2007qj,ArkaniHamed:2008yf}
and the double copy \cite{Kawai:1985xq,Berends:1988zp,Bern:2010ue,Bern:2019prr}
are most successful. At their core, they are connected by a simple
fact: cubic vertices are enough to describe any tree level graviton
amplitude. In BCFW we see this via onshell recursions. In the double
copy, graviton amplitudes are recast as two copies of gluon amplitudes
with a special trivalent configuration using the color-kinematic duality
\cite{Bern:2008qj}. Indeed, pure graviton amplitudes have been 
shown to be recursively described by a cubic action with auxiliary
fields that is classically equivalent to the Einstein-Hilbert action
\cite{Cheung:2017kzx}. This \emph{strictification} has been formally
demonstrated in \cite{Nutzi:2018vkl} using $L_{\infty}$ algebras.

Our universe, on the other hand, is not pure gravity. Our interest
resides in the study of interactions between gravitons and matter
particles. In this case, results using BCFW recursions (e.g. \cite{Cheung:2008dn,Falkowski:2020aso}),
double copy (e.g. \cite{Johansson:2019dnu}) or other diagrammatic
techniques (\cite{Plefka:2018zwm}), are considerably scarce, subject
to different subtleties/limitations that have so far eluded a more
systematic and practical output. At the dawn of gravitational waves
detection and black hole observation, any advance in the understanding
and formulation of the scattering of gravitons by matter is very welcome.
This letter is a step in this direction.

To our avail, the tree level information of a given field theory
can be elegantly extracted from its classical equations of motion
\cite{Boulware:1968zz}. This idea was further explored in \cite{Bardeen:1995gk,Cangemi:1996rx},
and later streamlined by the perturbiner method \cite{Rosly:1996vr,Rosly:1997ap,Selivanov:1998hn}.
As it turns out, there is an inspired multiparticle ansatz for the
solution of classical equations of motion that can be used to define
an offshell recursion for tree level amplitudes in terms of Berends-Giele
currents \cite{Berends:1987me}. This method recently regained interest
\cite{Mafra:2015gia,Lee:2015upy,Mafra:2015vca} and has been since
explored in different contexts \cite{Mafra:2016ltu,Mafra:2016mcc,Mizera:2018jbh,Lopez-Arcos:2019hvg,Guillen:2021mwp}.
Rather surprisingly, perturbiner methods have never been fully applied
to gravity, except for the very early analysis of the self-dual case
in \cite{Rosly:1997ap} and a simplified version for conformal supergravity
amplitudes in \cite{Johansson:2018ues}. Naively, a proper recursive
solution cannot be defined in a theory with an infinite number of
vertices. As we will show, however, there is a way around this obstacle
in gravity. 

In this work we propose a series of multiparticle solutions to Einstein's
field equations based on the perturbiner method. These solutions encompass
a broad class of interesting cases and can be applied to any two-derivative
matter field theory coupled to gravity. We can then define $n$-point
tree level scattering amplitudes between gravitons and matter particles
using a similar prescription to the super Yang-Mills case \cite{Berends:1987me}.
In this prescription, diffeomorphism invariance is manifest, with
a clear decoupling of pure gauge states. In addition, the analysis
of the soft limit behavior at leading order is surprisingly transparent.
First, we discuss pure gravity with a subsequent coupling to bosonic
matter. We then recast the Einstein-Hilbert action in terms of the
vielbein and the spin connection, in order to introduce the coupling
to fermionic matter and, consequently, supersymmetry. Our results
are agnostic to the number of spacetime dimensions and can be easily
automated. Whether or not there is an underlying worldsheet description,
they provide a compact and efficient computation of the scattering
of gravitons and matter at tree level.

\section{Field equations and gravitons}

Einstein's field equations without cosmological constant can be cast
as
\begin{equation}
R_{\mu\nu}-\tfrac{1}{2}g_{\mu\nu}R=\kappa T_{\mu\nu}.\label{eq:EFE}
\end{equation}
The right-hand side is the matter energy-momentum tensor, $T_{\mu\nu}$,
multiplied by the gravitational constant $\kappa$. On the left-hand
side, $g_{\mu\nu}$ denotes the spacetime metric (with inverse $g^{\mu\nu}$),
$R=g^{\mu\nu}R_{\mu\nu}$ is the scalar curvature and $R_{\mu\nu}$
is the Ricci tensor. As usual, $R_{\mu\nu}\equiv R_{\phantom{\sigma}\mu\rho\nu}^{\rho}$,
where $R_{\phantom{\sigma}\mu\nu\rho}^{\sigma}$ is the Riemann tensor,
\begin{equation}
R_{\phantom{\sigma}\mu\nu\rho}^{\sigma}=\partial_{\nu}\Gamma_{\mu\rho}^{\sigma}-\partial_{\rho}\Gamma_{\mu\nu}^{\sigma}+\Gamma_{\nu\lambda}^{\sigma}\Gamma_{\mu\rho}^{\lambda}-\Gamma_{\rho\lambda}^{\sigma}\Gamma_{\mu\nu}^{\lambda},\label{eq:Riemann}
\end{equation}
and $\Gamma_{\mu\nu}^{\sigma}=g^{\rho\sigma}\Gamma_{\mu\nu\rho}$
is the Christoffel symbol, with
\begin{equation}
\Gamma_{\mu\nu\rho}=\tfrac{1}{2}(\partial_{\mu}g_{\nu\rho}+\partial_{\nu}g_{\mu\rho}-\partial_{\rho}g_{\mu\nu}).\label{eq:Christoffel}
\end{equation}
The field equations (\ref{eq:EFE}) are covariant under general coordinate
transformations ($\delta x^{\mu}=\lambda^{\mu}$), with the metric
transforming as
\begin{equation}
\delta g_{\mu\nu}=g_{\mu\rho}\partial_{\nu}\lambda^{\rho}+g_{\nu\rho}\partial_{\mu}\lambda^{\rho}+\lambda^{\rho}\partial_{\rho}g_{\mu\nu}.\label{eq:generalgauge}
\end{equation}

In the absence of matter, (\ref{eq:EFE}) reduces to
\begin{equation}
R_{\mu\nu}=0,\label{eq:EHeom}
\end{equation}
which can be used to analyze linearized solutions around a given background,
i.e. the gravitons. These single-particle solutions around flat space
(with metric $\eta_{\mu\nu}$) are given by
\begin{equation}
g_{\mu\nu}(x)=\eta_{\mu\nu}+h_{\mu\nu}\boldsymbol{e}^{ik\cdot x},\label{eq:graviton-wave}
\end{equation}
with $k\cdot x=k_{\mu}x^{\mu}$. The graviton polarization, $h_{\mu\nu}$,
satisfies $\eta^{\nu\rho}k_{\rho}h_{\mu\nu}=\eta^{\mu\nu}h_{\mu\nu}=0$.
There is also a residual gauge transformation of the form $\delta h_{\mu\nu}=k_{\mu}\lambda_{\nu}+k_{\nu}\lambda_{\mu}$,
with $k\cdot\lambda=0$.

\section{Multiparticle solutions and recursions}

We can now look at the multiparticle solutions of the graviton field,
$g_{\mu\nu}(x)$, satisfying (\ref{eq:EHeom}). Consider

\begin{eqnarray}
g_{\mu\nu}(x) & = & \eta_{\mu\nu}+\sum_{P}H_{P\mu\nu}\boldsymbol{e}^{ik_{P}\cdot x},\label{eq:g-multiparticle}
\end{eqnarray}
where $H_{P\mu\nu}$ represents the multiparticle currents. The word
$P$ denotes a sequence of ordered letters, $P=p_{1}\ldots p_{n}$,
where $p_{i}$ is a single-particle label, with $k_{P}\equiv k_{p_{1}}+\ldots+k_{p_{n}}$.

In order to find the solutions for $H_{P\mu\nu}$, we have also to
work with $g^{\mu\nu}(x)$ satisfying $g^{\mu\rho}g_{\rho\nu}=\delta_{\nu}^{\mu}.$
For the expansion
\begin{eqnarray}
g^{\mu\nu}(x) & = & \eta^{\mu\nu}-\sum_{P}I_{P}^{\mu\nu}\boldsymbol{e}^{ik_{P}\cdot x},\label{eq:g-inverse}
\end{eqnarray}
the inverse identity implies that the currents $I_{P}^{\mu\nu}$ are
constrained to be
\begin{equation}
I_{P}^{\mu\nu}=\eta^{\mu\rho}\eta^{\nu\sigma}H_{P\rho\sigma}-\eta^{\nu\sigma}\sum_{P=Q\cup R}I_{Q}^{\mu\rho}H_{R\rho\sigma},\label{eq:I-recursion}
\end{equation}
where the sum goes over all deshuffles of $P$ into ordered words
$Q$, $R$ (see e.g. \cite{Mizera:2018jbh}). Although not explicitly,
$I_{P}^{\mu\nu}=I_{P}^{\nu\mu}$ and this can be recursively demonstrated
order by order in the sub-deshuffles.

Multiparticle currents with one-letter words are simply associated
to their single-particle equivalents (polarizations): $H_{p\mu\nu}=h_{p\mu\nu}$ and $I_{p}^{\mu\nu}= \eta^{\mu \rho} \eta^{\nu \sigma} h_{p\rho\sigma}$.

To every $x$-dependent object we will associate a multiparticle expansion.
For example, the Christoffel symbol can be expressed as $\Gamma_{\mu\nu\rho}=\sum_{P}\Gamma_{P\mu\nu\rho}\boldsymbol{e}^{ik_{P}\cdot x}$,
with
\begin{equation}
\Gamma_{P\mu\nu\rho}\equiv\tfrac{i}{2}(k_{P\mu}H_{P\nu\rho}+k_{P\nu}H_{P\mu\rho}-k_{P\rho}H_{P\mu\nu}).
\end{equation}
The parameter of general coordinate transformations may too be cast
as a multiparticle expansion as
\begin{equation}
\lambda^{\mu}=-i\sum_{P}\Lambda_{P}^{\mu}\boldsymbol{e}^{ik_{P}\cdot x}.\label{eq:gauge-multiparticle}
\end{equation}
This way, the gauge transformation (\ref{eq:generalgauge}) implies
that
\begin{multline}
\negthickspace\negthickspace\negthickspace\delta H{}_{P\mu\nu}\negthickspace=\negthickspace\negthickspace\sum_{P=Q\cup R}\negthickspace\Lambda_{Q}^{\rho}\{k_{Q\mu}H{}_{R\nu\rho}+k_{Q\nu}H_{R\mu\rho}+k_{R\rho}H{}_{R\mu\nu}\}\\
+k_{P\mu}\Lambda_{P\nu}+k_{P\nu}\Lambda_{P\mu}.\label{eq:multiparticle-transformation}
\end{multline}

We will choose the gauge $\eta^{\mu\nu}\Gamma_{\mu\nu\rho}=0$. This
is simpler than the de Donder gauge $g^{\mu\nu}\Gamma_{\mu\nu}^{\rho}=0$,
because its multiparticle version does not involve deshuffles, being
neatly expressed as
\begin{equation}
\eta^{\mu\nu}\Gamma_{P\mu\nu\rho}=i\eta^{\mu\nu}(k_{P\mu}H_{P\nu\rho}-\tfrac{1}{2}k_{P\rho}H_{P\mu\nu})=0.\label{eq:harmonic-gauge}
\end{equation}

In this gauge, the multiparticle currents of the Ricci tensor, $\mathcal{R}_{P\mu\nu}$,
are computed to be
\begin{multline}
\negthickspace\negthickspace\negthickspace\mathcal{R}_{P\mu\nu}=\tfrac{s_{P}}{2}H_{P\mu\nu}-i\negthickspace\negthickspace\sum_{P=Q\cup R}\negthickspace\negthickspace I_{Q}^{\rho\sigma}(k_{P\rho}\Gamma_{R\mu\nu\sigma}-k_{P\nu}\Gamma_{R\mu\rho\sigma})\\
-\eta^{\alpha\beta}\eta^{\rho\sigma}\negthickspace\negthickspace\sum_{P=Q\cup R}\negthickspace\negthickspace(\Gamma_{Q\nu\alpha\sigma}\Gamma_{R\mu\rho\beta}-\Gamma_{Q\rho\alpha\sigma}\Gamma_{R\mu\nu\beta})\\
+\eta^{\alpha\beta}\negthickspace\negthickspace\sum_{P=Q\cup R\cup S}\negthickspace\negthickspace I_{Q}^{\rho\sigma}(\Gamma_{R\nu\rho\beta}\Gamma_{S\mu\alpha\sigma}-\Gamma_{R\alpha\rho\beta}\Gamma_{S\mu\nu\sigma})\\
+\eta^{\alpha\beta}\negthickspace\negthickspace\sum_{P=Q\cup R\cup S}\negthickspace\negthickspace I_{Q}^{\rho\sigma}(\Gamma_{R\nu\alpha\sigma}\Gamma_{S\mu\rho\beta}-\Gamma_{R\rho\alpha\sigma}\Gamma_{S\mu\nu\beta})\\
-\negthickspace\negthickspace\sum_{P=Q\cup R\cup S\cup T}\negthickspace\negthickspace I_{Q}^{\rho\sigma}I_{R}^{\alpha\beta}(\Gamma_{S\nu\alpha\sigma}\Gamma_{T\mu\rho\beta}-\Gamma_{S\rho\alpha\sigma}\Gamma_{T\mu\nu\beta})\label{eq:Ricci-currents}
\end{multline}
where $s_{P}\equiv\eta^{\mu\nu}k_{P\mu}k_{P\nu}$ denotes the generalized
Mandelstam variables. The recursion relation for $H_{P\mu\nu}$ is
then obtained using equation (\ref{eq:EHeom}), i.e. $\mathcal{R}_{P\mu\nu}=0$.

\section{Tree level amplitudes}

Motivated by the Berends-Giele prescription \cite{Berends:1987me},
the tree level amplitude for the scattering of $n$ gravitons is defined
as
\begin{eqnarray}
\mathcal{M}_{n} & \equiv & \kappa\lim_{s_{2...n}\to0}s_{2...n}\,h_{1\mu\nu}I_{2...n}^{\mu\nu},\nonumber \\
 & = & \kappa\lim_{s_{2...n}\to0}s_{2...n}\,h_{1}^{\mu\nu}H_{2...n\mu\nu},\label{eq:n-graviton-amplitude}
\end{eqnarray}
on the support of momentum conservation. Whenever convenient, we will
raise or lower spacetime indices using the flat metric.

By construction, $H_{P\mu\nu}$ is symmetric in the exchange of any
two single-particle labels. This symmetry is lifted to the amplitude
$\mathcal{M}_{n}$, which is also symmetric in the exchange of any
two graviton legs, although only $(n-1)$ are manifest through $H_{2...n\mu\nu}$.
The particle in the first leg can be thought of as an off-shell leg
in the multiparticle recursion, then placed onshell in the definition
of the amplitude in (\ref{eq:n-graviton-amplitude}) via momentum
conservation and the limit $s_{2...n}\to0$.

The amplitude $\mathcal{M}_{n}$ is invariant under the residual transformations
of the graviton polarizations described after equation (\ref{eq:graviton-wave}).
In order to see this, we can examine the residual gauge transformations
preserving (\ref{eq:harmonic-gauge}). They lead to a recursion for
the currents $\Lambda_{P\mu}$ in (\ref{eq:gauge-multiparticle}),
given by
\begin{multline}
\Lambda_{P\mu}=-\tfrac{k_{P}^{\nu}}{s_{P}}\sum_{P=Q\cup R}\Lambda_{Q}^{\rho}(k_{Q\mu}H{}_{R\nu\rho}\\
+k_{Q\nu}H_{R\mu\rho}+k_{R\rho}H{}_{R\mu\nu}).\label{eq:residual-transformation}
\end{multline}
It is then just an algebraic step to show the invariance of $\mathcal{M}_{n}$
under (\ref{eq:multiparticle-transformation}) with multiparticle
parameters (\ref{eq:residual-transformation}).

The three-point amplitude is given by the well-known result
\begin{equation}
\mathcal{M}_{3}(h_{1},h_{2},h_{3})=\tfrac{\kappa}{4}h_{\mu\nu}^{1}h_{\alpha\beta}^{2}h_{\gamma\delta}^{3}V^{\mu\alpha\gamma}V^{\nu\beta\delta},
\end{equation}
in terms of the three-point Yang-Mills vertex
\[
V^{\mu\alpha\gamma}=(k_{2}^{\mu}-k_{3}^{\mu})\eta^{\alpha\gamma}+(k_{3}^{\alpha}-k_{1}^{\alpha})\eta^{\mu\gamma}+(k_{1}^{\gamma}-k_{2}^{\gamma})\eta^{\mu\alpha}.
\]

The current $H_{P\mu\nu}$ effectively describes interactions with
vertices from three to five points, as can be seen from the number
of deshuffles in (\ref{eq:Ricci-currents}). The four- or higher-point
amplitudes will not be explicitly displayed here, as their size grows
rapidly due to the nested deshuffles. We found it easier to perform
most of the cross-checks numerically, since it is straightforward
to implement the recursions for $H_{P\mu\nu}$ computationally.

\subsection*{Soft limit}

The soft limit of graviton amplitudes has a universal behavior \cite{Weinberg:1964ew,Weinberg:1965nx},
constituting a natural test for our proposal in (\ref{eq:n-graviton-amplitude}).
As it turns out, its soft limit analysis is very simple at leading
order.

We will take $h_{\mu\nu}^{1}$ as the soft graviton and parametrize
its momentum as $k_{1}^{\mu}=\tau q^{\mu}$, with $q^{2}=0$ and parameter
$\tau\to0$. In the soft limit, we can directly identify the dominant
contributions in $H_{23..n\mu\nu}$, for they come from the poles
of the generalized Mandelstam variables with $(n-2)$ momenta. For
example,
\begin{equation}
s_{3...n}=(\tau q+k_{2})^{2}=2\tau(q\cdot k_{2}),
\end{equation}
is attached to the multiparticle current with $(n-2)$ particles,
$H_{3..n\rho\sigma}$. We can then reexamine the recursion of the  $(n-1)$--particle
currents $H_{23..n\mu\nu}$ and readily express it as
\begin{multline}
s_{23..n}H_{23..n\mu\nu}=k_{2\mu}k_{2\nu}h_{2}^{\rho\sigma}H_{3..n\rho\sigma}\\
+\textrm{sym}(2,3,...,n)+\mathcal{O}(\tau^{0}),
\end{multline}
where $\textrm{sym}(2,3,...,n)$ takes care of the symmetrization
of the single-particle labels.

In terms of the amplitude, this parametrization leads to the leading
order contribution
\begin{equation}
\lim_{\tau\to0}\mathcal{M}_{n}=\tfrac{1}{\tau}\Big(\sum_{a=2}^{n}\tfrac{k_{a\mu}h_{1}^{\mu\nu}k_{a\nu}}{2(q\cdot k_{a})}\Big)\mathcal{M}_{n-1}(h_{2},...,h_{n}),\label{eq:soft-graviton}
\end{equation}
manifesting the universal Weinberg pole. Since diffeomorphism invariance
is inbuilt in our results, subleading soft limits should be directly
reproduced \cite{Cachazo:2014fwa,Chakrabarti:2017ltl}.

\section{Matter coupled to gravity}

The matter contributions to (\ref{eq:EFE}) come from the energy-momentum
tensor
\begin{equation}
T_{\mu\nu}\equiv-2\frac{1}{\sqrt{-g}}\frac{\delta}{\delta g^{\mu\nu}}S_{\textrm{matter}},\label{eq:Tmatter}
\end{equation}
where $S_{\textrm{matter}}$ is the matter action. In terms of a multiparticle
expansion, we have $T_{\mu\nu}=\sum_{P}\mathcal{T}_{P\mu\nu}\boldsymbol{e}^{ik_{P}\cdot x}$,
where the form of the currents $\mathcal{T}_{P\mu\nu}$ is particular
to the model. In order for this to make sense, the single-particle
solutions of the free equations of motion associated to the matter
action must be described in terms of plane waves. These are our asymptotic
states. 

The recursion relations for the currents $H_{P\mu\nu}$ are obtained
by plugging the corresponding multiparticle expansions in (\ref{eq:EFE}).
The result is
\begin{eqnarray}
\mathcal{R}_{P\mu\nu} & = & \tfrac{1}{2}\eta_{\mu\nu}\eta^{\rho\sigma}\mathcal{R}_{P\rho\sigma}+\kappa\mathcal{T}_{P\mu\nu}\nonumber \\
 &  & +\tfrac{1}{2}\sum_{P=Q\cup R}(H_{Q\mu\nu}\eta^{\rho\sigma}-\eta_{\mu\nu}I_{Q}^{\rho\sigma})\mathcal{R}_{R\rho\sigma}\nonumber \\
 &  & -\tfrac{1}{2}\negthickspace\sum_{P=Q\cup R\cup S}\negthickspace H_{Q\mu\nu}I_{R}^{\rho\sigma}\mathcal{R}_{S\rho\sigma},\label{eq:g-recursion-matter}
\end{eqnarray}
where $\mathcal{R}_{P\mu\nu}$ is defined in (\ref{eq:Ricci-currents}).
Naturally, we recover $\mathcal{R}_{P\mu\nu}=0$ when $\mathcal{T}_{P\mu\nu}=0$.

The amplitude prescription is the same of (\ref{eq:n-graviton-amplitude}),
but now we are able to describe the scattering of matter bosons and
gravitons.

\subsection{Massive scalar}

Our first example is the massive scalar coupled to gravity and otherwise
free, with equation of motion
\begin{equation}
(g^{\mu\nu}\partial_{\mu}\partial_{\nu}-m^{2})\phi=g^{\mu\nu}\Gamma_{\mu\nu}^{\rho}\partial_{\rho}\phi,\label{eq:eom-scalar}
\end{equation}
and covariantly conserved energy-momentum tensor
\begin{equation}
T_{\mu\nu}=-\partial_{\mu}\phi\partial_{\nu}\phi+\tfrac{1}{2}g_{\mu\nu}(g^{\rho\sigma}\partial_{\rho}\phi\partial_{\sigma}\phi+m^{2}\phi^{2}).\label{eq:Tmn-scalar}
\end{equation}

Equation (\ref{eq:eom-scalar}) leads to the following recursion for
the scalar multiparticle currents, $\Phi_{P}$, 
\begin{multline}
\negthickspace\negthickspace(s_{P}+m^{2})\Phi_{P}=\negthickspace\sum_{P=Q\cup R}(I_{Q}^{\mu\nu}k_{R\mu}k_{R\nu}+\eta^{\mu\nu}k_{R}^{\rho}\Gamma_{Q\mu\nu\rho})\Phi_{R}\\
-\sum_{P=Q\cup R\cup S}\eta^{\mu\nu}I_{Q}^{\rho\sigma}\Phi_{S}(\Gamma_{R\mu\nu\sigma}k_{S\rho}+\Gamma_{R\rho\sigma\nu}k_{S\mu})\\
+\sum_{P=Q\cup R\cup S\cup T}I_{Q}^{\mu\nu}I_{R}^{\rho\sigma}\Gamma_{S\mu\nu\sigma}k_{T\rho}\Phi_{T}.
\end{multline}
Similarly, equation (\ref{eq:Tmn-scalar}) leads to
\begin{multline}
\negthickspace\negthickspace\negthickspace\mathcal{T}_{P\mu\nu}=\sum_{P=Q\cup R}\{k_{Q\mu}k_{R\nu}+\tfrac{1}{2}\eta_{\mu\nu}[m^{2}-(k_{Q}\cdot k_{R})]\}\Phi_{Q}\Phi_{R}\\
+\tfrac{1}{2}\sum_{P=Q\cup R\cup S}H_{Q\mu\nu}\Phi_{R}\Phi_{S}[m^{2}-(k_{Q}\cdot k_{R})]\\
+\tfrac{1}{2}\sum_{P=Q\cup R\cup S}\eta_{\mu\nu}I_{Q}^{\rho\sigma}k_{R\rho}k_{S\sigma}\Phi_{R}\Phi_{S}\\
+\tfrac{1}{2}\sum_{P=Q\cup R\cup S\cup T}H_{Q\mu\nu}I_{R}^{\rho\sigma}k_{S\rho}k_{T\sigma}\Phi_{S}\Phi_{T}.
\end{multline}

These quantities are then used to compute the tree level scattering
of gravitons and massive scalars. For example, the four point amplitude
with two gravitons ($h_{1}$, $h_{2}$) and two massive scalars ($3$,
$4$) is given by
\begin{multline}
\mathcal{M}_{4}=2\kappa^{2}(k_{2}\cdot h_{1}\cdot h_{2}\cdot k_{1})-\tfrac{1}{2}\kappa^{2}s_{34}(h_{1}\cdot h_{2})\\
+2\kappa^{2}[(k_{3}\cdot h_{1}\cdot h_{2}\cdot k_{4})+(k_{4}\cdot h_{1}\cdot h_{2}\cdot k_{3})]\\
+\tfrac{4\kappa^{2}}{s_{34}}\big\{\tfrac{1}{2}(h_{1}\cdot h_{2})[(k_{1}\cdot k_{3})(k_{2}\cdot k_{4})+(k_{1}\cdot k_{4})(k_{2}\cdot k_{3})]\\
+(k_{2}\cdot h_{1}\cdot k_{3})(k_{4}\cdot h_{2}\cdot k_{1})+(k_{2}\cdot h_{1}\cdot k_{4})(k_{3}\cdot h_{2}\cdot k_{1})\\
-(k_{3}\cdot h_{1}\cdot k_{4})(k_{1}\cdot h_{2}\cdot k_{1})-(k_{3}\cdot h_{2}\cdot k_{4})(k_{2}\cdot h_{1}\cdot k_{2})\\
-(k_{2}\cdot h_{1}\cdot h_{2}\cdot k_{3})(k_{1}\cdot k_{4})-(k_{2}\cdot h_{1}\cdot h_{2}\cdot k_{4})(k_{1}\cdot k_{3})\\
-(k_{3}\cdot h_{1}\cdot h_{2}\cdot k_{1})(k_{2}\cdot k_{4})-(k_{4}\cdot h_{1}\cdot h_{2}\cdot k_{1})(k_{2}\cdot k_{3})\big\}\\
+\tfrac{4\kappa^{2}}{(s_{23}+m^{2})}(k_{4}\cdot h_{1}\cdot k_{4})(k_{3}\cdot h_{2}\cdot k_{3})\\
+\tfrac{4\kappa^{2}}{(s_{24}+m^{2})}(k_{3}\cdot h_{1}\cdot k_{3})(k_{4}\cdot h_{2}\cdot k_{4}),
\end{multline}
matching known results in the literature, e.g. \cite{Holstein:2008sx,Bjerrum-Bohr:2019nws}.

\subsection{Yang-Mills theory}

Here we provide the ingredients for computing the scattering of gravitons
and gauge vectors. 

The energy-momentum tensor of $S_{\textrm{YM}}$ is given by
\begin{equation}
T_{\mu\nu}=\tfrac{2}{g_{\textrm{YM}}^{2}}\textrm{Tr}(g^{\rho\sigma}F_{\mu\rho}F_{\nu\sigma}-\tfrac{1}{8}g_{\mu\nu}g^{\lambda\rho}g^{\delta\sigma}F_{\lambda\delta}F_{\rho\sigma}),\label{eq:Tmn-YM}
\end{equation}
where $g_{\textrm{YM}}$ is the coupling constant, $F_{\mu\nu}=\partial_{\mu}A_{\nu}-\partial_{\nu}A_{\mu}-i[A_{\mu},A_{\nu}]$
is the field-strength, and the trace $\textrm{Tr}$ is taken over
a given non-Abelian group or just $U(1)$ for Maxwell's theory. The
equations of motion of the gauge field can be cast as
\begin{equation}
g^{\nu\rho}D_{\nu}F_{\mu\rho}=g^{\nu\rho}[A_{\nu},F_{\mu\rho}],\label{eq:eom-YM}
\end{equation}
where $D_{\mu}$ denotes the curved space covariant derivative,
\begin{equation}
D_{\nu}F_{\mu\rho}=\partial_{\nu}F_{\mu\rho}-\Gamma_{\nu\mu}^{\sigma}F_{\alpha\rho}-\Gamma_{\nu\rho}^{\sigma}F_{\mu\alpha}.
\end{equation}

It is then straightforward to plug the multiparticle expansion $A_{\mu}=\sum_{P}\mathcal{A}_{P\mu}\boldsymbol{e}^{ik_{P}\cdot x}$
back in (\ref{eq:eom-YM}) and obtain a recursive definition for the
currents $\mathcal{A}_{P\mu}$. The covariant gauge $g^{\mu\nu}D_{\mu}A_{\nu}=0$
seems to be the simplest choice in this case.

These Einstein-Yang-Mills amplitudes can then be compared with other
results in the literature obtained through different techniques, e.g.
\cite{Fu:2017uzt,Chiodaroli:2017ngp,Teng:2017tbo,Du:2017gnh}.

\section{Fermions and Supersymmetry}

In order to consider fermions, we turn the local Lorentz group into
a gauge symmetry. The spacetime metric $g_{\mu\nu}$ is mapped to
the (local) flat metric, $\eta_{ab}$, using the vielbein, $e_{\mu}^{a}$
(with inverse $e_{a}^{\mu}$), such that $g_{\mu\nu}=\eta_{ab}e_{\mu}^{a}e_{\nu}^{b}.$
The gauge field of the Lorentz symmetry is the spin connection, $\omega_{\mu}^{ab}$,
and the \emph{flattened} Riemann tensor, $R_{\mu\nu}^{ab}\equiv\eta^{ac}e_{\sigma}^{b}e_{c}^{\rho}R_{\phantom{\sigma}\rho\mu\nu}^{\sigma}$,
can be seen as its field strength, given by
\begin{equation}
R_{\mu\nu}^{ab}\equiv\partial_{\mu}\omega_{\nu}^{ab}+\eta_{cd}\omega_{\mu}^{ac}\omega_{\nu}^{db}-(\mu\leftrightarrow\nu),\label{eq:flat-Riemann}
\end{equation}
with scalar curvature $R=e_{a}^{\mu}e_{b}^{\nu}R_{\mu\nu}^{ab}$.

Spinor couplings to the curved background are implemented by replacing
spacetime derivatives by their Lorentz-covariant version. Given a
spinor $\psi$, its covariant derivative is defined as
\begin{equation}
D_{\mu}\psi=\partial_{\mu}\psi+\tfrac{1}{4}\omega_{\mu}^{ab}\Gamma_{ab}\psi,
\end{equation}
where $\Gamma_{ab}=\tfrac{1}{2}[\Gamma_{a},\Gamma_{b}]$ and $\Gamma_{a}$
denotes the usual gamma matrices satisfying $\{\Gamma_{a},\Gamma_{b}\}=2\eta_{ab}$.

Next, we rewrite Einstein's field equations in terms of the vielbein
and, independently, the spin connection. This is known as the Palatini
variation. In the presence of matter, they take the form
\begin{eqnarray}
R_{\mu}^{a} & = & \kappa T_{\mu}^{a}+\tfrac{1}{2}e_{\mu}^{a}R,\\
\omega_{\mu}^{ab} & = & \kappa W_{\mu}^{ab}+\tfrac{1}{2}e^{\nu[a}(\partial_{\mu}e_{\nu}^{b]}-\Gamma_{\mu\nu\rho}e^{\rho b]}),\label{eq:eom-spinconnecton}
\end{eqnarray}
where $R_{\mu}^{a}\equiv e_{b}^{\nu}R_{\mu\nu}^{ab}$, $[ab]=ab-ba$,
and $\Gamma_{\mu\nu\rho}$ is the zero-torsion Christoffel symbols
in (\ref{eq:Christoffel}). The matter tensors are defined as
\begin{eqnarray}
T_{\mu}^{a} & \equiv & e\frac{\delta}{\delta e_{a}^{\mu}}S_{matter},\\
W_{\mu}^{ab} & \equiv & eP_{\mu\nu}^{ab,cd}\frac{\delta}{\delta\omega_{\nu}^{cd}}S_{matter},
\end{eqnarray}
with $e=\det e_{a}^{\mu}$, and
\begin{equation}
P_{\mu\nu}^{ab,cd}\equiv\eta^{d[a}(e_{\nu}^{b]}e_{\mu}^{c}+\tfrac{1}{2}\eta^{b]c}g_{\mu\nu}+\tfrac{2}{(D-2)}e_{\mu}^{b]}e_{\nu}^{c}).
\end{equation}

From here onward, the perturbiner method goes as usual. We define
the multiparticle expansion for the vielbein and its inverse analogously
to the metric expansions in (\ref{eq:g-multiparticle}) and (\ref{eq:g-inverse}),
\begin{eqnarray}
e_{\mu}^{a} & = & \delta_{\mu}^{a}+\sum_{P}E_{P\mu}^{a}\boldsymbol{e}^{ik_{P}\cdot x},\label{eq:E-multiparticle}\\
e_{a}^{\mu} & = & \delta_{a}^{\mu}-\sum_{P}F_{P\nu}^{b}\boldsymbol{e}^{ik_{P}\cdot x}.\label{eq:F-multiparticle}
\end{eqnarray}
The \emph{mixed} Kronecker deltas $\delta_{\mu}^{a}$ and $\delta_{a}^{\mu}$
indicate that the vielbeins are expanded around \emph{flat space}.
The inverse relations $e_{\mu}^{a}e_{a}^{\nu}=\delta_{\mu}^{\nu}$
and $e_{\mu}^{a}e_{b}^{\nu}=\delta_{b}^{a}$ constrain $F_{Pa}^{\mu}$
to satisfy
\begin{eqnarray}
F_{Pa}^{\mu} & = & \delta_{a}^{\nu}\delta_{b}^{\mu}E_{P\nu}^{b}-\delta_{a}^{\nu}\sum_{P=Q\cup R}E_{Q\nu}^{b}F_{Rb}^{\mu},\nonumber \\
 & = & \delta_{a}^{\nu}\delta_{b}^{\mu}E_{P\nu}^{b}-\delta_{b}^{\mu}\sum_{P=Q\cup R}E_{Q\nu}^{b}F_{Ra}^{\nu}.\label{eq:FrecursiveE}
\end{eqnarray}
The proof of equivalence between the first and second lines follows
the same logic of $I_{P}^{\mu\nu}=I_{P}^{\nu\mu}$ after equation
(\ref{eq:I-recursion}).

We can then use general coordinate transformations and local Lorentz
symmetry to fix a convenient gauge. We found the simplest one to be\begin{subequations}\label{eq:vielbein-gauge}
\begin{eqnarray}
(\eta^{\mu\nu}\eta_{ab}-\tfrac{1}{2}\delta_{a}^{\mu}\delta_{b}^{\nu})\partial_{\mu}e_{\nu}^{b} & = & 0,\\
\delta_{a}^{\mu}e_{\mu b}-\delta_{b}^{\mu}e_{\mu a} & = & 0.
\end{eqnarray}
\end{subequations}The first equation is a truncated version of (\ref{eq:harmonic-gauge}),
while the second is known as symmetric gauge. In terms of the multiparticle
currents, this gauge has a simple realization and does not involve
deshuffles. The single-particle polarizations, $e_{p\mu}^{a}$, satisfy
$k_{p}^{\mu}e_{p\mu}^{a}=\delta_{a}^{\mu}e_{p\mu}^{a}=0$, with residual
gauge symmetry
\begin{equation}
\delta e_{p\mu a}=\delta_{a}^{\nu}(k_{p\mu}\lambda_{p\nu}+k_{p\nu}\lambda_{p\mu})\label{eq:residual-gauge-vielbein}
\end{equation}
and $k_{p}\cdot\lambda_{p}=0$.

With these choices, the recursion for $E_{P\mu}^{a}$ can be written
as\begin{widetext}
\begin{multline}
s_{P}E_{P\mu}^{a}=\kappa(\delta_{a}^{b}\delta_{\mu}^{\nu}+\tfrac{1}{(2-D)}\delta_{\mu}^{a}\delta_{b}^{\nu})\mathcal{T}_{P\nu}^{b}-i\kappa\delta_{b}^{\nu}(k_{P\mu}\mathcal{W}_{P\nu}^{ab}-k_{P\nu}\mathcal{W}_{P\mu}^{ab})\\
+\tfrac{1}{2}\delta_{b}^{\nu}\sum_{P=Q\cup R}[k_{P\nu}F_{Q}^{\rho a}(k_{R\mu}E_{R\rho}^{b}-k_{R\rho}E_{R\mu}^{b})-k_{P\mu}F_{Q}^{\rho a}(k_{R\nu}E_{R\rho}^{b}-k_{R\rho}E_{R\nu}^{b})-(a\leftrightarrow b)]\\
+\sum_{P=Q\cup R}[iF_{Qb}^{\nu}(k_{R\mu}\Omega_{R\nu}^{ab}-k_{R\nu}\Omega_{R\mu}^{ab})-\eta_{cd}\delta_{b}^{\nu}(\Omega_{Q\mu}^{ac}\Omega_{R\nu}^{db}-\Omega_{Q\nu}^{ac}\Omega_{R\mu}^{db})+\tfrac{\kappa}{(2-D)}(E_{Q\mu}^{a}\delta_{b}^{\nu}-\delta_{\mu}^{a}F_{Qb}^{\nu})\mathcal{T}_{R\nu}^{b}]\\
+\tfrac{1}{2}\delta_{b}^{\nu}\sum_{P=Q\cup R}[\eta^{\rho a}\eta^{\sigma b}(k_{P\mu}E_{Q\nu}^{c}-k_{P\nu}E_{Q\mu}^{c})-(k_{P\mu}\delta_{\nu}^{c}-k_{P\nu}\delta_{\mu}^{c})(F_{Q}^{\rho a}\eta^{\sigma b}-F_{Q}^{\rho b}\eta^{\sigma a})](k_{R\sigma}E_{R\rho c}-k_{R\rho}E_{R\sigma c})\\
+\tfrac{1}{2}\delta_{b}^{\nu}\sum_{P=Q\cup R\cup S}[(k_{P\mu}\delta_{\nu}^{c}-k_{P\nu}\delta_{\mu}^{c})F_{Q}^{\rho a}F_{R}^{\sigma b}-(k_{P\mu}E_{Q\nu}^{c}-k_{P\nu}E_{Q\mu}^{c})(F_{R}^{\rho a}\eta^{\sigma b}-F_{R}^{\rho b}\eta^{\sigma a})](k_{S\sigma}E_{S\rho c}-k_{S\rho}E_{S\sigma c})\\
+\sum_{P=Q\cup R\cup S}[\eta_{cd}F_{Qb}^{\nu}(\Omega_{R\mu}^{ac}\Omega_{S\nu}^{db}-\Omega_{R\nu}^{ac}\Omega_{S\mu}^{db})-\tfrac{\kappa}{(2-D)}E_{Q\mu}^{a}F_{Rb}^{\nu}\mathcal{T}_{S\nu}^{b}]\\
+\tfrac{1}{2}\delta_{b}^{\nu}\sum_{P=Q\cup R\cup S\cup T}(k_{P\mu}E_{Q\nu}^{c}-k_{P\nu}E_{Q\mu}^{c})F_{R}^{\rho a}F_{S}^{\sigma b}(k_{T\sigma}E_{T\rho c}-k_{T\rho}E_{T\sigma c}),
\end{multline}
\end{widetext}where we have used the multiparticle currents of $T_{\mu}^{a}$,
$W_{\mu}^{ab}$ and $\omega_{\mu}^{ab}$, respectively $\mathcal{T}_{P\mu}^{a}$,
$\mathcal{W}_{P\mu}^{ab}$ and $\Omega_{P\mu}^{ab}$.

The $n$-point tree level amplitudes are defined as
\begin{eqnarray}
\mathcal{M}_{n} & \equiv & \kappa\lim_{s_{2...n}\to0}s_{2...n}\,e_{1a}^{\mu}E_{2...n\mu}^{a}.\label{eq:vielbein-amplitude}
\end{eqnarray}
Similarly to (\ref{eq:n-graviton-amplitude}), the invariance of this
amplitude under the residual gauge transformations (\ref{eq:residual-gauge-vielbein})
has a straightforward demonstration in the gauge (\ref{eq:vielbein-gauge}). 

Now that we are able to consistently account for fermionic degrees
of freedom in the multiparticle solutions, it is just a small step
to consider supersymmetric field theories, in particular supergravity.

\section{Final remarks}

In this letter we have found multiparticle solutions to Einstein's
field equations, with a compact recursive definition for graviton
multiparticle currents in $D$-dimensional Minkowski space. These
currents can then be used to compute any tree level scattering between
gravitons and matter particles, with or without supersymmetry.

The key insight is the recursive definition of the inverse metric
$g^{\mu\nu}$ in (\ref{eq:g-inverse})-(\ref{eq:I-recursion}), with
an analogous expression for the inverse vielbein $e_{a}^{\mu}$ in
(\ref{eq:F-multiparticle})-(\ref{eq:FrecursiveE}). Effectively,
this recursion works as a truncation of the gravity action. It is
yet another way of seeing that the infinite number of graviton vertices,
though required by diffeomorphism invariance, play no role at the
tree level dynamics.

The practical appeal of our formulas is that they can be easily computerized.
For pure gravity amplitudes, this might not present an advantage over
current methods, in particular the double copy construction using
color-kinematics duality \cite{Bern:2010ue,Bern:2019prr}, and BCFW
on-shell recursion \cite{Bedford:2005yy,Cachazo:2005ca}. Nevertheless,
for the mixed scattering of gravitons and matter particles, the ingredients
presented here constitute a versatile tool for computing tree level
amplitudes in a broad class of theories. Our results do not require
an underlying worldsheet theory (a \emph{stringy} origin) and can
be applied to more general field configurations. We believe our solutions
can become a robust standard for such amplitudes, with a reliable
framework to be used to test both (1) the extension of current methods
and (2) possible new techniques for tree level scattering.

Towards a more efficient algorithm implementation, the recursions
we presented can be recast in a ``color-stripped'' form. The name
is inherited from the Yang-Mills perturbiner, where such a construction
is more natural due to the color structure. For the graviton, a color-stripped
perturbiner is not limited by the ordered words in the multiparticle
expansion. In this case, the word splitting needed to recursively
define the currents is greatly simplified (see e.g. \cite{Mizera:2018jbh}).
For example, the color-dressed deshuffle $P=Q\cup R$ of a $n$-letter
word leads to $(2^{n}-2)$ pairs of ordered words. This is to be compared
with mere $(n-1)$ pairs of the deconcatenation $P=QR$ in the color-stripped
form. We just have to be careful to properly symmetrize the final
amplitudes with respect to the graviton legs, but this is computationally
much less costly. 

A more immediate extension of our results would be to consider multiparticle
expansions in curved space. While this cannot be efficiently developed
in general backgrounds, we found that the perturbiner extension to
(anti) de Sitter spaces leads to an intuitive recursive definition
of Witten diagrams \cite{Witten:1998qj} for different matter fields.
This is being explored in an ongoing project \cite{GLL:2021}.

Finally, a quick comment on loop computations. The perturbiner method
seems to be intrinsically classical: it consists of solving equations
of motion. However, recent results using homotopy algebras in quantum
field theory \cite{Jurco:2020yyu} have uncovered that loop-level
scattering amplitudes can also be recursively computed. The natural
question then is: can we reformulate these recursions as solutions
of some \emph{quantum equation of motion}? Although this speculation
looks a bit far-fetched, based on very preliminary investigations
we think the answer might be affirmative.
\begin{acknowledgments}
We would like to thank Subhroneel Chakrabarti, Arthur Lipstein, Carlos
Mafra and Oliver Schlotterer for valuable comments and reference suggestions.
HG is supported by the Royal Society via a PDRA grant. RLJ acknowledges
the Czech Science Foundation - GA\v{C}R for financial support under
the grant 19-06342Y.
\end{acknowledgments}

\end{document}